\documentclass[useAMS,usenatbib]{mn2e}
\usepackage{aas_macros}
\usepackage{graphicx,amssymb,natbib,textcomp,fontenc,lscape,color,afterpage}
\usepackage{deluxetable}
\newcommand{\fv}{F_\mathrm{var}}

\title[Millimeter-band variability of NGC~7469]
{Millimeter-band variability of the radio-quiet nucleus of NGC 7469}
\author[Baldi et al.]
{Ranieri D. Baldi$^{1}$\thanks{E-mail: baldi@ph.technion.ac.il}, Ehud Behar$^{1}$, Ari Laor$^{1}$, Assaf Horesh$^{2}$
\\
$^{1}$Department of Physics, Technion 32000, Haifa 32000, Israel\\
$^{2}$Weizmann Institute of Science, Rehovot, Israel}
\begin{document}



\maketitle

\label{firstpage}

\begin{abstract}

We report short-cadence monitoring of a radio-quiet (RQ) Active
Galactic Nuclei (AGN), NGC\,7469, at 95\,GHz (3\,mm) over a period of
70 days with the CARMA telescope.  The AGN varies significantly
($\pm3\sigma$ from the mean) by a factor of two within 4-5 days.  The
intrinsic 95\,GHz variability amplitude in excess of the measurement
noise (10\%) and relative to the mean flux is comparable to that in
the X-rays, and much higher than at 8.4\,GHz.  The mm-band variability
and its similarity to the X-ray variability adds to the evidence that
the mm and X-ray emission have the same physical origin, and are
associated with the accretion disk corona.

\end{abstract}


\section{Introduction}

Radio and mm-wave observations provide crucial information for our
understating of accretion mechanisms and particle acceleration
processes in Active Galactic Nuclei (AGN). The nature of accelerators
in radio-loud (RL) AGN is known to be manifested by the collimated
relativistic jets, launched from the innermost region close to the
black hole accretion disk, and can extend up to Mpc scales.
Conversely, for radio-quiet (RQ) AGN the origin of the radio emission
is still unclear.  In the past it was claimed that the
presence or lack of relativistic jets is the reason for the RL/RQ
dichotomy (e.g. \citealt{kellermann94}). The presence of pc-kpc scale
  extended radio emission in RQ AGN, which could be due to
  sub-relativistic velocities, complicates this distinction
  (e.g. \citealt{blundell96,nagar99,nagar00,ulvestad05_n4151}).
  Whether there is a radio bimodality, or whether the radio luminosity
  distribution is continuous is still under debate
  (e.g. \citealt{white00}).

 Different origins have been proposed to account for the radio properties of RQ AGN: 
 synchrotron radiation from down-scaled jets \citep{barvainis96,gallimore06,orienti10}; 
 thermal emission/absorption from hot gas \citep{gallimore04};
 synchrotron emission from an optically thick advection-dominated
 accretion flow \citep{doi11};
 and coronal emission from magnetic activity above the accretion disk \citep{field93}. 

 \citet{laor08} investigated the origin of the radio emission in RQ
 AGN using the Palomar quasar sample.  They demonstrated that their
 radio (at 5 GHz) and X-ray (0.2- 20~keV) luminosities are correlated and
 that they follow the well established correlation $L_R = 10^{-5}L_X$
 for coronally active cool stars \citep{gudel93}.  This correlation
 over 20 orders of magnitude in luminosity suggests perhaps that the
 radio emission in RQ AGN is also related to coronal, magnetic
 activity similar to that of stars. In phenomena of magnetic
 reconnection, coronal mass ejections may be the origin for the
 extended radio emission in RQ AGN.

Radio monitoring of AGN in the cm band in the past decades mostly focused on RL AGN,
 and specifically on Blazars (e.g., 
 \citealt{raiteri08b,lister09}).
Conversely, radio monitoring of RQ AGN is rare. 
Detected variability of RQ AGN is typically $<$50\% at 5-15 GHz on a year time scale
\citep{neff83,wrobel00,mundell09,jones11}.
\citet{falcke01} found that RQ AGN are more variable than RL AGN expect for Blazars,
and that low-luminosity RQ AGN are among the most variable radio sources,
likely due to the small black-hole mass/size.

The mm-band holds several advantages over the cm-band.
First, mm-waves can probe a smaller region of emission due to the improving angular resolution with frequency
for a given array.
Moreover, synchrotron self absorption decreases with frequency.
The size, therefore, of an opaque synchrotron source 
at 95\,GHz is about $\sim$20 times smaller than at 5\,GHz, and could vary 20 times faster.
The rapid variability provides insight into the AGN core not accessible even to interferometric imaging. 
If the variability is related to the X-ray variability, it could indicate that the
two emitting regions are spatially coincident.

The mm spectrum of AGN typically bridges the high radio frequency synchrotron emission from a jet/outflow 
and the low-frequency IR thermal dust emission. 
RQ AGN show either flat or inverted spectra at these high radio frequencies, suggesting
optically-thick core emission \citep{doi05,doi11}.
We recently found a 95\,GHz excess of a factor of up to $\sim$7 with respect to the 
observed low-frequency steep spectra, for a sample of X-ray
bright RQ AGN \citep{behar15}.


Little data are available on mm-band monitoring of RQ AGN. Worthy campaigns have been performed for
the low-luminosity M~81 \citep{sakamoto01,schodel07} and Centaurus A \citep{israel08}, 
which vary by a factor of a few on a day timescale.
Sgr\,A$^{*}$ also features intraday variability of $\sim$40\% in mm wavelengths
\citep{miyazaki04,lu11,dexter14}.
\citet{doi11} found significant 3-mm variability within months of low-luminosity RQ AGN.

In this paper, we report results from a dedicated monitoring program
of a RQ AGN with the CARMA telescope (the Combined Array for
Research in Millimeter-wave Astronomy).  We monitored NGC\,7469 at 3-mm
(95 GHz) over 70 days with periods of daily monitoring.  As far as we
know, this is the first time a RQ AGN was monitored so persistently at
mm wavelengths.  The paper is organized as follows. In
Sect.~\ref{sec:sample} we describe the observations.  In
Sect.~\ref{sec:results} we present the light curve and quantify the
variability.  In Sect.~\ref{sec:summary} we discuss the results and
draw conclusions.

\begin{table}
 \centering
\caption{Observation log}
\begin{tabular}{cccc|cc}
  \hline
Grade & Date  &  time &  Flux calib   & F$_{peak}$ & F$_{tot}$    \\
 \multicolumn{6}{c}{{\bf Array C}} \\
\hline
  A-  &  14MAR17   & 2.20 & Neptune  &   3.11$\pm$0.30 & 7.73  \\
  A-  &  14MAR18   & 2.22 & MWC349   &   2.29$\pm$0.24 & 7.44  \\
  A   &  14MAR21   & 2.42 & MWC349   &   2.49$\pm$0.25 & 6.78  \\    
  B+  &  14MAR25   & 2.27 & Neptune  &   2.88$\pm$0.28 & 7.89  \\
  C+  &  14MAR27   & 2.40 & Neptune  &   3.18$\pm$0.33 & 8.49  \\
  A-  &  14MAR28   & 2.38 & Neptune  &   2.54$\pm$0.26 & 6.84  \\
  C+  &  14MAR29   & 2.56 & Neptune  &   2.43$\pm$0.27 & 8.54  \\
  A-  &  14MAR31   & 2.39 & Neptune  &   3.11$\pm$0.30 & 7.73  \\
  A-  &  14APR01   & 2.19 & Neptune  &   3.82$\pm$0.36 & 10.29  \\
  C-  &  14APR05   & 3.65 & Neptune  &   2.74$\pm$0.32 & 7.96  \\
  B+  &  14APR07   & 3.67 & Neptune  &   3.42$\pm$0.28 & 8.43  \\
  A-  &  14APR11   & 2.42 & Neptune  &   4.06$\pm$0.36 & 8.57  \\
  C-  &  14APR13   & 1.42 & Neptune  &   3.31$\pm$0.44 & 10.69  \\
  B+  &  14APR15   & 2.24 & MWC349   &   3.04$\pm$0.32 & 7.64  \\
  A-  &  14APR17   & 2.59 & Neptune  &   3.08$\pm$0.35 & 8.20   \\
  B   &  14APR18   & 1.95 & Neptune  &   3.05$\pm$0.41 & 7.18  \\
  C+  &  14APR20   & 2.32 & Neptune  &   2.17$\pm$0.33 & 7.64  \\
  A-  &  14APR21   & 2.35 & Neptune  &   3.27$\pm$0.35 & 8.05  \\
\hline
 \multicolumn{6}{c}{{\bf Array D}} \\
  B+  &  14APR27   & 2.34  & Neptune &   6.44$\pm$0.85 & 13.01  \\
  A-  &  14APR28   & 2.37  & Neptune &   6.80$\pm$0.80 & 12.47  \\
  B-  &  14MAY06   & 2.86  & Neptune &   6.63$\pm$0.49 & 14.82  \\ 
  B-  &  14MAY20   & 2.83  & Neptune &   5.96$\pm$0.43 & 12.62  \\
  A-  &  14MAY25   & 1.39  & Neptune &   6.35$\pm$0.43 & 13.83  \\
  \hline
\end{tabular}
\label{obslog}
Column description: (1)  dataset grade, (2) observation
  date, (3) total observation time
  on NGC~7469 in hours, (4) flux calibrator, (5) peak flux from the
  gaussian fit in mJy/beam for NGC~7469, (6) total integrated
  flux in mJy for the NGC~7469.
\end{table}

\section{Observations}
\label{sec:sample}


NGC\,7469 ($z=0.01588, D_L=71.2$\,Mpc) was drawn from a recent AGN sample selected based on the high X-ray brightness and documented variability \citep{behar15}.
It was chosen for intense monitoring, because it is one of the brightest targets in the sample. 
Its 95~GHz flux density is in excess ($\times1.5$) over the steep spectrum at low-radio frequencies \citep[see][]{behar15}.
\citet{pereztorres09} reported seven 8.4-GHz observations of NGC\,7469 over 8 years and its radio core (0\farcs5) flux density was consistently between 12-13 mJy, i.e., no significant variability.

We observed continuum emission in the 3 mm (95~GHz) window from the
nucleus of NGC\,7469 with the CARMA telescope. CARMA is a 15 element
interferometer consisting of nine 6.1 meter antennas and six 10.4
meter antennas, located in California (USA).  We monitored NGC\,7469
over a period of 70 days with a cadence ranging from 1 day to a few
days.  We obtained a total of 38 flux measurements, 29 and 9, in the
C- and D- array configuration respectively, reaching an angular
resolution of 2\farcs2, and 5\farcs5, respectively.

\begin{figure*}
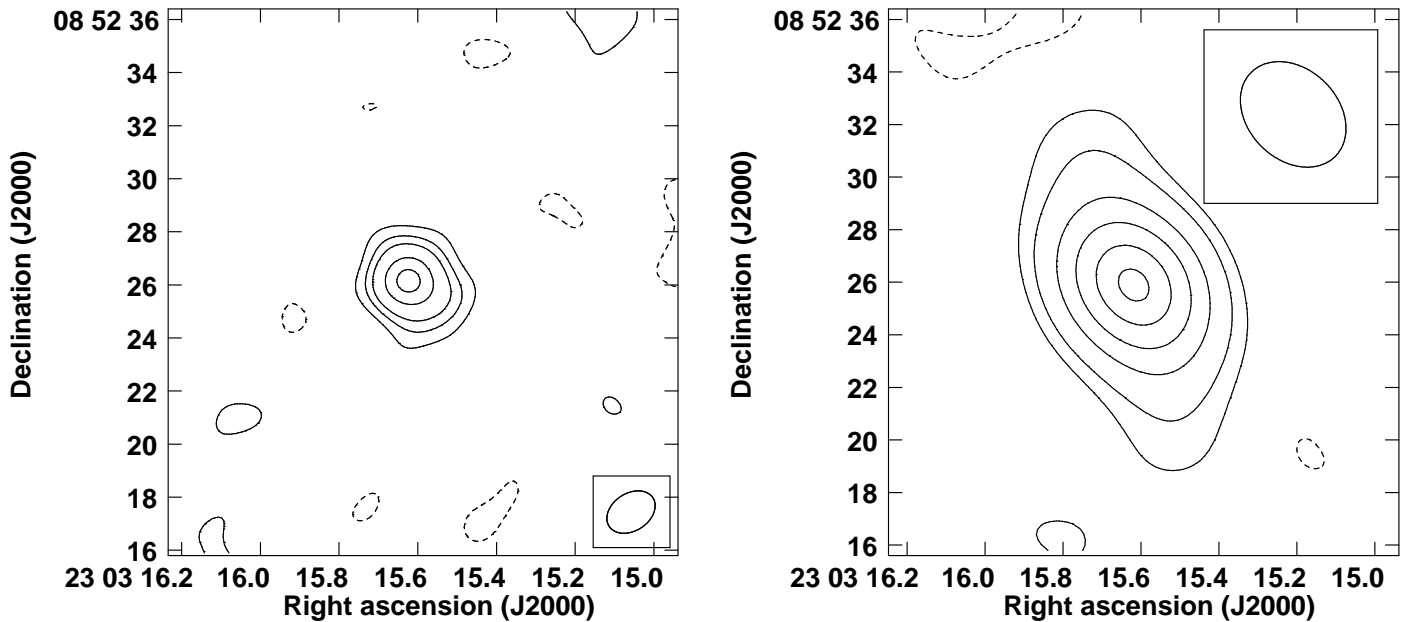

\centerline{
\hskip -1cm
\includegraphics[scale=0.45,angle=-90]{NGC7469.1C_cont.ps}
\includegraphics[scale=0.45,angle=-90]{NGC7469.5D_cont.ps}}
\caption{Two examples of radio maps of NGC~7469 at 95 GHz obtained
  with C- (left panel) and D-array (right panel) configurations. In
  the box there is the beam size of the map.  
  1\arcsec\ in NGC\,7469 corresponds to 0.35\,kpc.
  The contour levels of
  the maps are respectively: left panel (-0.45, 0.45, 0.90, 1.35,
  2.25, 2.93) mJy/beam and right panel (-0.56, 0.70, 1.40, 2.80, 4.20,
  5.60, 6.65) mJy/beam.}
\label{ngc7469map}
\end{figure*}

Only observations with a weather grade of C or better
(Table~\ref{obslog}) are used, in order to ensure low noise in the
final maps.  Furthermore, some observations were not usable due to
telescope internal failures, and to target pointing that was too
short. Effectively, 23 observations (18+5 in the two configurations)
of NGC\,7469 are used in our analysis and listed in Table~\ref{obslog}. 
Absolute flux calibrators were used at the beginning of each observation track,
and are also listed in Table~\ref{obslog}.
  
The MIRIAD software package \citep{sault95} was used to reduce the
visibility data, including flagging data affected by shadowed
antennas, poor weather or antenna malfunctions.  The observing
sequence was to integrate on a phase calibrator for $\sim$2 minutes
and on the primary target for 15 minutes. This cycle was repeated
10-12 times for a total of $\sim$4 hours for each observation. The
passband and gain calibrator was 3C\,454.3.  We also observed a
secondary target, 2218-035, a RL quasar, for comparison to make sure
the variability of NGC\,7469 is genuine.  We performed
self-calibration of the data, which turned out to be unnecessary, as
it changed the flux by not more than 7\%.  The amplitude solutions
show a constant pattern across the baselines ranging between
$\sim$0-110 kilo$\lambda$, which is a sign of the presence of a
point-source.  The phase solutions are stable in the range of
$\pm20^{\circ}$ independent of the baselines. The phase and amplitude
solutions were applied to the targets using standard procedures. The
maps were obtained by inverting and cleaning with a natural weight of
the visibility data (robustness parameter = 2). Different robustness
parameters (in the standard range of [-2,2]) produce final maps
differing by not more than 5\% in the point-source flux.  Finally,
using the MIRIAD {\it imfit} package we measured the flux using a
point-source gaussian fit. The observation log and flux density
measurements are given in Table~\ref{obslog}.



\section{RESULTS}
\label{sec:results}


The 95\,GHz maps of NGC~7469 
with 2\farcs2 and 5\farcs5 beams are shown in Fig.~\ref{ngc7469map}.  
There is a hint of a slight elongation of the AGN point source towards the NW,
which is reminiscent of the radio extension observed also at 5 and 8.5 GHz on a scale of
$\sim$10\arcsec\ in the same direction \citep{condon80,ulvestad81,kukula95}. 
We perform a gaussian fit to the source and
measure the peak flux, which corresponds to the point-source flux
(mJy/beam), and then we measure the total integrated flux for the extended emission.
The mean peak flux is 3.0 and 6.4 mJy, respectively for the C and D arrays.
This is typically 1/3, and 1/2 of the total integrated flux, in the two arrays (Table~\ref{obslog}). 
This means that part of the peak flux includes unresolved extended emission, 
which will dilute the genuine core variability.

\begin{figure*}
\centerline{
\hskip -1cm
\includegraphics[scale=0.4,angle=90]{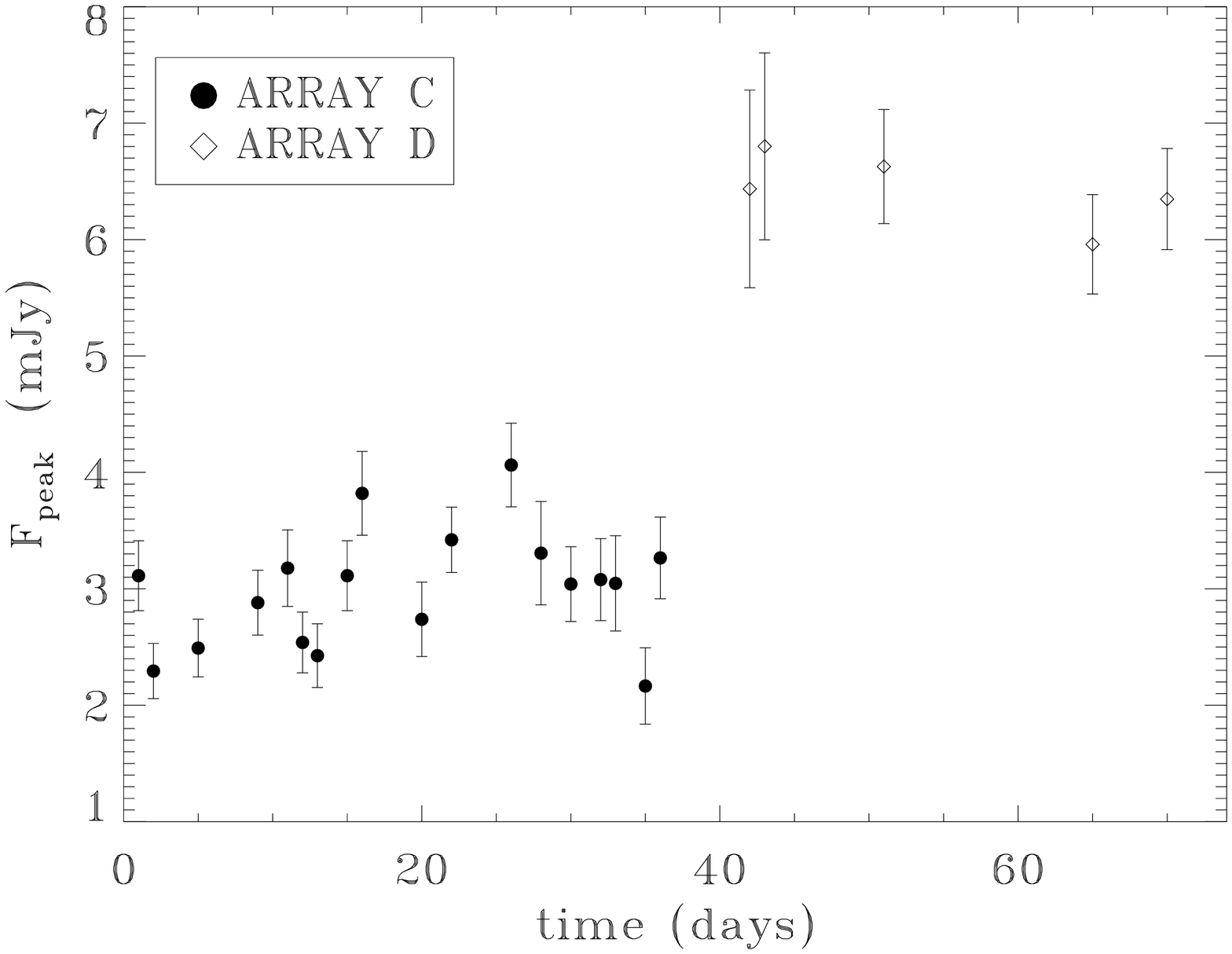}
\includegraphics[scale=0.4,angle=90]{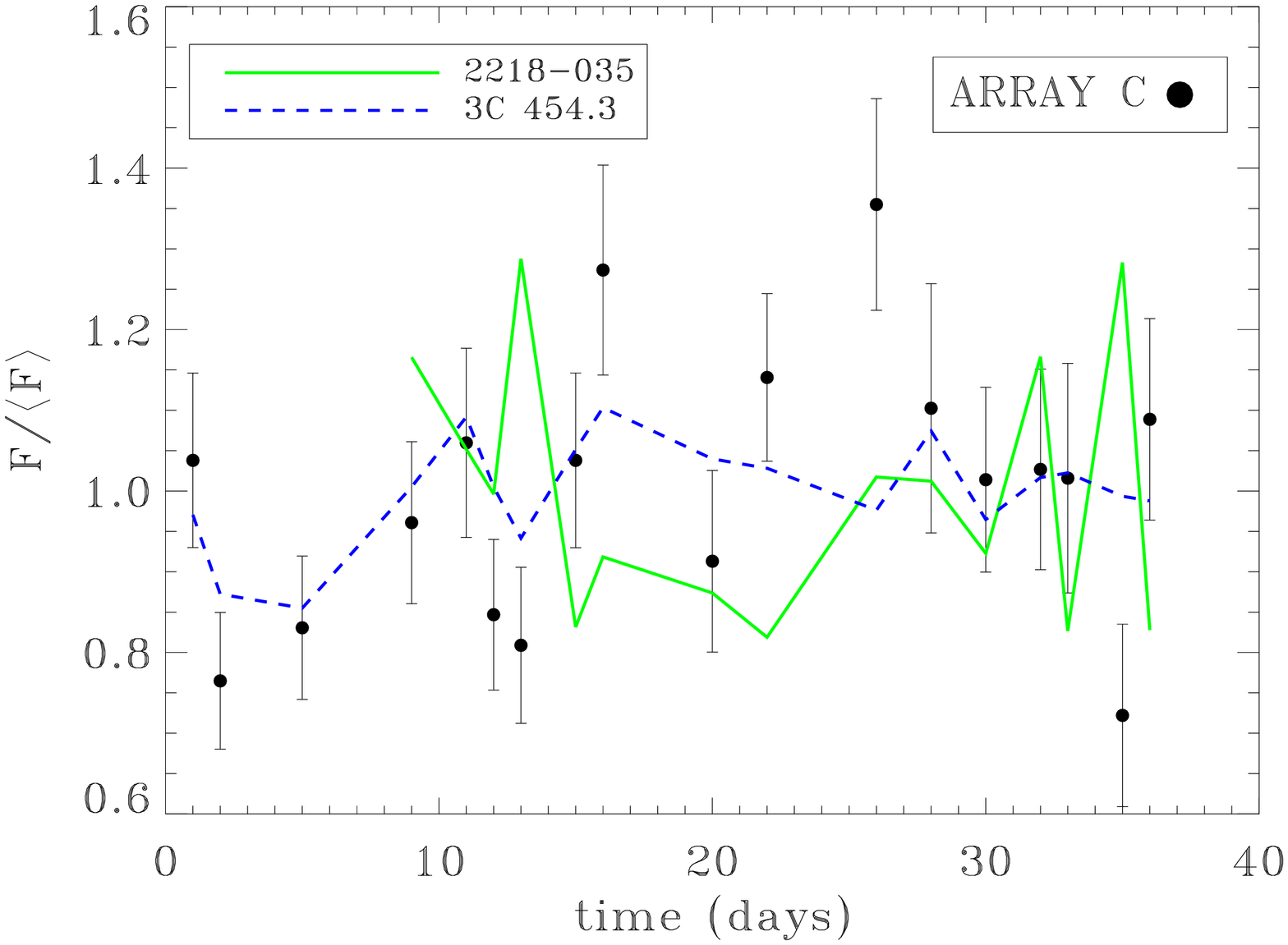}}
\caption{Peak flux light curves at 95 GHz. Left panel: NGC\,7469 with the C and D arrays.
Right panel: Flux to mean ratio using the C-array data only: for NGC\,7469, the bandpass calibrator 3C\,454.3, 
and the secondary target 2218-035 (not observed in the first three epochs). Errors on the last two targets are larger than those of NGC\,7469, and are not plotted.
}
\label{ngc7469lc}
\end{figure*}

The 95 GHz light curve of NGC\,7469 is shown in the left panel of
Figure~\ref{ngc7469lc}.  The peak flux ranges between 2$-$4~mJy in the C-array, and between 6$-$7~mJy in D 
(Table~\ref{obslog}), which correpsond to luminosity range, 39.06$-$39.61 erg s$^{-1}$. 
The factor of $\sim$2 difference in the two configuration fluxes is due to the different beam sizes.  
Hence, we hereafter treat the fluxes obtained with the two arrays separately.
The right panel of Figure~\ref{ngc7469lc} shows the peak-flux light curve normalized to the average for the C array only.
Variability of up to a factor of 2 ($\pm$30\% amplitude) is observed. 
The variability amplitude of the total flux (not shown) is only 20\% that of the peak flux, 
as the core gets diluted by extended, steady emission. 
For the same reason, no variability can be detected in the D array.

In the right panel of Fig.~\ref{ngc7469lc}, 
the blue dashed line represents the flux-to-mean ratio of the bandpass calibrator, 3C\,454.3, 
which is a flat-spectrum radio quasar known to vary on year/month scale \citep{villata09}. 
Indeed, its fluctuations in Fig.~2 are only 10\% those of NGC\,7469.
The green solid line represents the secondary target, 2218-035. 
The variability of 3C\,454.3 and of 2218-035 is much less significant than that of NGC\,7469, as quantified in the following section. 
More importantly, their trends are generally different from how NGC\,7469 varies, 
which lends to the authenticity of the variability of NGC\,7469.

Interestingly, the strongest peak flux variability by a factor of 2 occurs within 4-5 days.
This suggests a 95-GHz source size of no more than 4-5 light days, or $\sim $0.004~pc,
which is still $\sim$10 times larger than the minimal physical size estimated from the luminosity of a self absorbed synchrotron source \citep[i.e., radio-sphere,][]{behar15}.

\subsection{Statistical tests of variability}

In order to quantify the 95-GHz variability pattern of NGC\,7469 we performed several tests, 
all of which confirm the source is significantly variable.

\subsubsection{$\fv$}
We calculate the Fractional variability $\fv$
\citep[e.g.][]{edelson02,vaughan03} of the 95~GHz light curve of NGC\,7469.
$\fv$ quantifies the intrinsic variability amplitude in excess of the measurement uncertainties,
and relative to the mean source flux density: 

\begin{equation}
\fv = \sqrt{S^2 - \langle \sigma ^2 \rangle \over \langle F_\nu
\rangle^2},
\end{equation}

\noindent where $S^2$ is the variance of the light curve, $\langle
\sigma^2 \rangle $ is the mean squared measurement uncertainty,
 and $ \langle F_\nu \rangle$ is the mean flux density (count rate is used in X-rays).
 The error on $\fv$ for $N$ flux measurements can be defined as

\begin{equation}
\sigma_{\fv} = \sqrt{  \left\{ \sqrt \frac{ \langle\sigma ^2\rangle}{N} \cdot \frac{1}{\langle F_\nu \rangle} \right\} ^2         +       \left\{ \sqrt \frac{1}{2N} \cdot \frac {  \langle\sigma ^2\rangle }{\langle F_\nu \rangle^2 \fv} \right\} ^2             }
\end{equation}


For the 18 data points in the C-array light curve measured over 36 days we obtain $\fv$(95\,GHz) = $12.8\%\pm$2.5\%. 
This is remarkably similar to $\fv$(2-12\,keV)= 16.0\%$\pm$0.4\% measured in the X-rays for NGC~7469 over 36 days by \citet{markowitz04}, 
using light curves from the RXTE telescope.
On shorter time scales of 6 and 1 days, they measured $\fv$(2-12\,keV) = 12.3\%$\pm$0.4\% and 7.5\%$\pm$1.0\%, respectively. 
In all cases, these values are a few percent less (more), when measured separately for the soft (hard) X-ray band. 
We measure $\fv$(95\,GHz) over 6 days to be $12.5\%\pm$6.3\%, which is consistent with the X-ray value. 
This value (and error) is a weighted mean of five 6-day periods during our campaign that had at least 4 observations each.
The small number of data points does not support calculating $\fv$ for the D-array. 

The $\fv$ estimator is similar to the 
De-biased Variability Index DVI \citep[ = $\fv / \sqrt{N}$,][]{akritas96}. For
the C-array observations $DVI = 3\%$.  This is consistent with the DVI
measured for most RQ quasars, i.e., DVI $\lesssim$10\% at 8.5\,GHz and
DVI $\lesssim$30\% at 15 GHz \citep{akritas96}.
We also measured $\fv$(95\,GHz) for the bandpass calibrator and the secondary target. 
Their $\fv$ values are consistent with 0\%, taking into account the measurement uncertainties,
which provides further evidence for the genuine variability of NGC\,7469.

\subsubsection{Fitting the u-v data}

\begin{figure}
\centerline{
\hskip -1cm
\includegraphics[scale=0.4,angle=90]{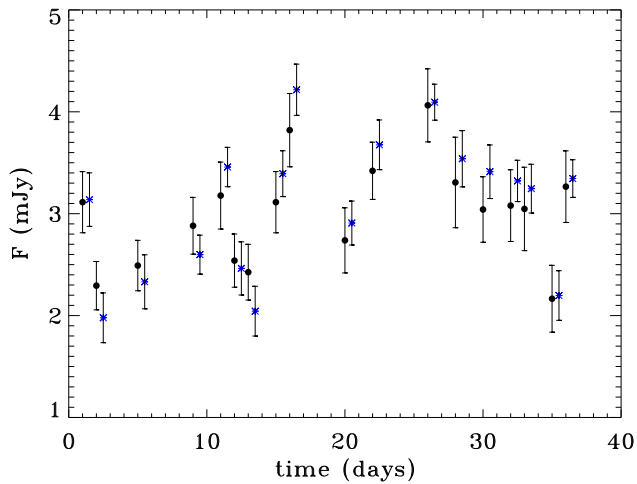}}
\caption{Light curve of NGC\,7469 (only array C). 
Black circles are  the flux density measurements from the cleaned maps, 
while blue stars (slightly shifted horizontally for clarity) are those obtained by directly fitting the uv visibility data.
Full consistency is apparent.}
\label{uvfit}
\end{figure}

In order to double check the robustness of the flux measurements and
variability we also fitted the visibility data in the C array for a
gaussian point source directly in the u-v plane, by using the
AIPS\footnote{The NRAO Astronomical Image Processing System (AIPS) is
  a package to support the reduction and analysis of data taken with
  radio telescopes.} {\it uvfit} task.  This fitting procedure tool
finds a component, unresolved at $\sim$2\arcsec, located at the
position of the core.  The flux densities of this component are
consistent with the values obtained from the cleaned maps
(Table~~\ref{obslog}) within 15\%.  This consistency is demonstrated
in Figure~\ref{uvfit}.  When the flux increases/decreases in the clean
images, it also increases/decreases in the u-v Gaussian fitting with
similar amplitudes.  Consequently, $\fv = 15.0\pm2.4\%$ in the
uv-fitted light curve, which is consistent with the value measured
from the cleaned maps.

We carried out one final test for the validity of the variability, and
its independence of the precise unresolved point-source structure.  We
repeated the {\it uvfit} gaussian fitting task, but now using only the
baselines larger than 50 kilo$\lambda$, which is half of the available
u-v range.  Since the long baselines probe small angular scales, this
procedure allows us to verify whether the emission is dominated by the
unresolved core.  The flux densities obtained are again consistent
with those measured from the cleaned maps to within 15\%. This
confirms that the compact core of NGC\,7469 dominates the fitted
gaussian flux, and is the variable component.  All these tests confirm
the genuine variability of NGC\,7469.

\subsubsection{$\chi ^2$}

Another variability estimator commonly used 
for gamma- and X- ray variability \citep{paolillo04,young12,lanzuisi14} is

\begin{equation} 
\chi^2 = \sum_{i=1}^{N} \frac{(F_{\nu, i} - \langle F_\nu \rangle)^2}{\sigma_i^2}, 
\end{equation}

\noindent where $F_{\nu, i}$ and $\sigma_i^2$ are each flux density measurement and
its associated uncertainty. $N$ and $\langle F_\nu \rangle$ are as before.
This $\chi ^2$ value tests how significantly the light curve differs from a constant. 
For an intrinsically non-variable source, the value of $\chi^2$ is
expected to be $N-1$ (i.e., $\chi^2/d.o.f=1$).  
The advantage of this estimator is that a  probability 
can be associated with the light curve being variable, i.e., rejection of the null-hypothesis.
$P > 95\%$ is considered significance evidence for variability.
The C-array light curve of Fig.~\ref{ngc7469lc} is variable at a confidence level of 99.9817\%.
Furthermore, we iteratively excluded each single data point and re-calculated $\chi ^2$ to make sure 
the value is not affected by a single (bad) point.
The confidence level for positive variability remains always $>95\%$,
confirming that NGC~7469 is significantly variable at 95~GHz.

\section{Summary and Conclusions}
\label{sec:summary}

The nucleus of NGC~7469 has been detected with the CARMA telescope at 95 GHz in 23 epochs 
over a period of 70 days, which is the most intensive short-cadence 
monitoring of a RQ AGN to date.
The 95\,GHz light curve of NGC~7469 is variable with an amplitude of up to $\pm$30\% (6$\sigma$ from the flux maximum to the minimum) from the mean. 
The source shows a rapid flux variation of a factor 2 in 4-5 days. 
If we consider the 95-GHz flux density of 5.0$\pm$0.2~mJy from \citet{behar15} observed with
the same CARMA (C) array in November 2013, the source varied by a factor of $\sim$1.7 in four months.
This variability amplitude of NGC\,7469 at 95 GHz is larger and on shorter time scales than that
reported at 8.4 GHz by \citet{pereztorres09} for the same AGN with annual monitoring over 8 years. 
The 95-GHz fractional variability amplitude $\fv$ is
consistent within the errors with the X-ray value and, moreover, the
time scale for variability in the two bands is similar, i.e., of the
order of 1~day. In conclusion, we statistically confirmed that NGC~7469
has a variable core at 95 GHz.

The variability and the radio-sphere size ($\sim10^{-4}$ pc) are
consistent with a compact core akin to the X-ray source. The 95-GHz to
X-ray (2-10 keV) luminosity ratio ($\sim10^{-4}$)\footnote{We use an
  X-ray (2-10 keV) luminosity L$_{X}$ = 1.5$\times$10$^{43}$ erg s$^{-1}$
  from \citet{shu10}.} is consistent with the \citeauthor{gudel93}
relation for cool stars, but much lower than for RL AGN. These
anlaogies point to the picture of a magnetically heated corona being
responsible for the 95 GHz emission for our object.  A model of a very
hot ($>10^{8}$ K) magnetically confined corona over the accretion
disk, similar to the observed structure of the solar corona was
postulated by \cite{galeev79}.  Phenomena of magnetic reconnection in
the corona would produce both X-ray and radio emission, as well as
coronal mass ejections, like in active stars, and could produce the
extended radio emission observed in RQ AGN.  The base for these
moderately accelerated and not well collimated outflows could be the
accretion disk corona in X-ray binaries and AGN
\citep{markoff05,liu14}.

The ultimate test of the coronal conjecture is simultaneous mm/X-ray
monitoring.  Clearly, the present study needs to be complemented and
further expanded with the high-sensitivity and high-resolution radio
telescopes, e.g. ALMA, IRAM, and later SKA.  Temporal correlation
between the X-ray and radio/mm light curves from RQ AGN, as the
Neupert effect seen in coronally active stars \citep{neupert68}, would
be the smoking gun of coronal radio emission.

\section*{Acknowledgments}

This research is supported by the I-CORE program of the Planning and
Budgeting Committee and the Israel Science Foundation (grant numbers
1937/12 and 1163/10), and by a grant from Israel's Ministry of Science
and Technology. RDB was supported at the Technion by a fellowship from
the Lady Davis Foundation. RDB thanks the referee for the comments
which help to confirm the genuine variability of the
target. We also thank the helpful discussion with C.~M. Raiteri,
A. Capetti, and I. McHardy.  E.B. received funding from the European
Unions Horizon 2020 research and innovation programme under the Marie
Sklodowska-Curie grant agreement No 655324.

This work was carried out with the CARMA telescope.  construction was
derived from the states of California, Illinois, and Maryland, the
James S. McDonnell Foundation, the Gordon and Betty Moore Foundation,
the Kenneth T. and Eileen L. Norris Foundation, the University of
Chicago, the Associates of the California Institute of Technology, and
the National Science Foundation.

\bibliography{my}

\end{document}